\documentclass[conference]{IEEEtran}
\IEEEoverridecommandlockouts
\usepackage{cite}
\usepackage{amsmath,amssymb,amsfonts}
\usepackage{graphicx}
\usepackage{textcomp}
\usepackage{xcolor}
\usepackage{booktabs}
\usepackage{multirow}
\usepackage{tabularx}
\usepackage{hyperref}
\hypersetup{
	colorlinks=true,
	linkcolor=blue,
	citecolor=black,
	filecolor=black,
	urlcolor=blue,
}
\graphicspath{ {images/} }
\def\BibTeX{{\rm B\kern-.05em{\sc i\kern-.025em b}\kern-.08em
    T\kern-.1667em\lower.7ex\hbox{E}\kern-.125emX}}
\begin{document}

\title{Phishing Detection Using Machine Learning Techniques \\
{\footnotesize \textsuperscript{}}
}

\author{\IEEEauthorblockN{Vahid Shahrivari}
\IEEEauthorblockA{{Computer Engineering Departmen} \\
{ Sharif University of Technology}\\
Tehran, Iran \\
v.shahrivari@student.sharif.edu}
\and
\IEEEauthorblockN{Mohammad Mahdi Darabi}
\IEEEauthorblockA{{School of Electrical and Computer Engineering} \\
{University of Tehran}\\
Tehran, Iran \\
mahdi.darabi@ut.ac.ir}
\and
\IEEEauthorblockN{Mohammad Izadi}
\IEEEauthorblockA{{Computer Engineering Departmen} \\
{ Sharif University of Technology}\\
Tehran, Iran \\
izadi@sharif.edu }
}

\maketitle

\begin{abstract}
The Internet has become an indispensable part of our life, However, It also has provided opportunities to anonymously perform malicious activities like Phishing. Phishers try to deceive their victims by social engineering or creating mock-up websites to steal information such as account ID, username, password from individuals and organizations. Although many methods have been proposed to detect phishing websites, Phishers have evolved their methods to escape from these detection methods. One of the most successful methods for detecting these malicious activities is Machine Learning. This is because most Phishing attacks have some common characteristics which can be identified by machine learning methods. In this paper, we compared the results of multiple machine learning methods for predicting phishing websites. 
\end{abstract}


\section{Introduction}
\begin{figure}[ht!]
	\centering
	\includegraphics[width=0.5\textwidth]{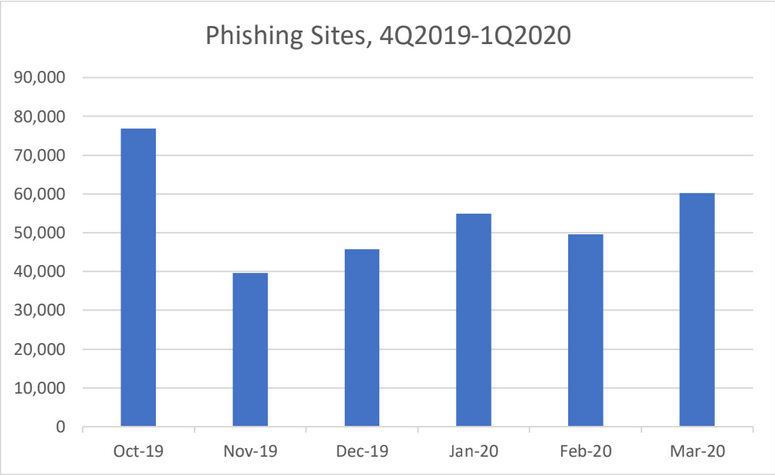}
	\caption{Total number of phishing websites detected by APWG \cite{APWG2019Q4}}
	\label{fig:fig1}
\end{figure}

Phishing is a kind of Cybercrime trying to obtain important or confidential information from users which is usually carried out by creating a counterfeit website that mimics a legitimate website. Phishing attacks employ a variety of techniques such as link manipulation, filter evasion, website forgery, covert redirect, and social engineering. The most common approach is to set up a spoofing web page that imitates a legitimate website. These type of attacks were top concerns in the latest 2018 Internet Crime Report, issued by the U.S. Federal Bureau of Investigation’s Internet Crime Complaint Center (IC3). The statistics gathered by the FBI’s IC3 for 2018 showed that internet-based theft, fraud, and exploitation remain pervasive and were responsible for a staggering \$2.7 billion in financial losses in 2018. In that year, the IC3 received 20,373 complaints against business email compromise (BEC) and email account compromise (EAC), with losses of more than \$1.2 billion \cite{FBIReport2018}. The report notes that the number of these sophisticated attacks have grown increasingly in recent years. Anti-Phishing Working Group(APWG) emphasizes that phishing attacks have grown in recent years, Figure \ref{fig:fig1} illustrates the total number of phishing sites detected by APWG in the first quarter of 2020 and the last quarter of 2019. This number has a gradual growth raising from 162,155 in the last quarter of 2019 to 165,772 cases in the first quarter of 2020. Phishing has caused severe damages to many organizations and the global economy, in the fourth quarter of 2019, APWG member OpSec Security found that SaaS and webmail sites remained the most frequent targets of phishing attacks. Phishers continue to harvest credentials from these targets by operating BEC and subsequently gain access to corporate SaaS accounts\cite{APWG2019Q4}.
Many approaches have been used to filter out phishing websites. Each of these methods is appliable on different stages of attack flow, for example, network-level protection, authentication, client-side tool, user education, server-side filters, and classifiers. Although there are some unique features in every type of phishing attack, most of these attacks depict some similarities and patterns. Since machine learning methods proved to be a powerful tool for detecting patterns in data, these methods have made it possible to detect some of the common phishing traits, therefore, recognizing phishing websites. In this paper, we provide a comparative and analytical evaluation of different machine learning methods on detecting the phishing websites. The machine learning methods that we studied are Logistic Regression, Decision Tree, Random Forest, Ada-Boost, Support Vector Machine, KNN, Artificial Neural Networks, Gradient Boosting, and XGBoost.
The rest of this paper is organized as follows:
in section \ref{section_1_1} we list some widely used phishing techniques, in Section \ref{section-2}
we discuss different types of phishing and phishing attack prevention methods. In section \ref{section-3} we provide an overview of different machine learning methods for phishing detection. In section \ref{section-4} we illustrate the features of our dataset. In section \ref{section-5} and \ref{section-6} we show evaluation results of suggested machine learning methods and finally we draw conclusions and discuss future works in section \ref{section-7}.

\section{Phishing techniques}\label{section_1_1}
In this section, we discuss some well-known phishing approaches used by criminals to deceive people.

\subsection{Link manipulation}
The phishing is mainly about links. There are some clever ways to manipulate a URL to make look like a legitimate URL. One method is to represent the malicious URLs as hyperlinks with name on websites. Another method is to use misspelled URLs which will look like a legitimate URL for example ghoogle.com. A variant of typosquatting that is much harder to recognize compared to mentioned link manipulation methods is called IDN Spoofing in which the attackers use a character in non-English language that looks exactly like an English character for example using a Cyrillic "c" or "a" instead of English counterparts\cite{link_manipulation}.

\subsection{Filter evasion}
Phishers show the content of their website in pictures or they use Adobe-Flash making it difficult to be detected by some phishing detection methods. To avoid this kind of attack using optical character recognition is required \cite{lam2009counteracting}.

\subsection{Website forgery}
In this type of attack, Phishing is happening at a legitimate website by manipulating the target website JavaScript code. These types of attacks which are also known as cross-site scripting are very hard to detect because the victim is using the legitimate website.

\subsection{Covert redirect}
This attacks targets websites using OAuth 2.0 and OpenID protocol. While trying to grant token access to a legitimate website, users are giving their token to a malicious service. However, this method did not gain much attention due to its low significance\cite{jing2017covert}.

\subsection{Social engineering}
This type of phishing is carried out through social interaction. It uses psychological tricks to deceive users to give away security information. This type of attack happens in multi-steps. At first, the phisher investigates the potential weak points of targets required for the attack. Then, the phisher tries to gain the target's trust and at last, provide a situation in which the target reveals important information. There are some social engineering phishing methods, namely, baiting, scareware, pretexting, and spear phishing\cite{krombholz2015advanced}.

\section{Phishing detection approaches: an overview}\label{section-2}
Various methods have been proposed to avert phishing attacks through each level of attack flow. Some of these methods require training the users to be prepared for future attacks and some of them work automatically and warn the user. These methods can be listed as follows:
\begin{itemize}
	\item User training
	\item Software detection
\end{itemize}

\subsection{User training}
Educating users and company employees and warning them about phishing attacks have an impact on preventing phishing attacks. Multiple methods have been proposed for training users. Many researches concluded that the most impactful approach to help the users to distinguish between phishing and legitimate websites is interactive teaching \cite{usertraining1} \cite{usertraining2}. Although user training is an effective method however humans errors still exist and people are prone to forget their training. Training also requires a significant amount of time and it is not much appreciated by non-technical users \cite{dhamija2006phishing}.

\subsection{Software detection}
Although user training can prevent some phishing attacks however we are bombarded every day by hundreds of websites therefore applying our training on each website is a cumbersome and sometimes non-practical task. Another alternative for detecting phishing websites is to use the software. The software can analyze multiple factors like the content of the website, email message, URL, and many other features before it makes its final decision which is more reliable than humans. Multiple software methods are proposed for phishing detection which is categorized as follows:
\begin{enumerate}
	\item \textit{List-base approach}: One of the widely used methods for phishing detection is using blacklist-based anti-phishing methods which are integrated into web browsers. These methods use two types of lists, namely the white list which contains the name of valid websites, and the blacklist which keeps the record of malicious websites. Usually, the blacklist is obtained either through user feedback or through third-party reports which are created by using another phishing detection scheme. Some studies have shown that blacklist-based anti-phishing approaches can detect 90 percent of the malicious website at the time of initial check \cite{blacklist1}.

\item \textit{Visual similarity-base approach}: One of the main reasons that people are tricked into believing that they are using a legitimate website but in reality, they are filling a form in a malicious website is that the phishing website appearance is exactly similar to the targeted legitimate website. Some methods use visual similarities by analyzing text content, text format, HTML, CSS, and images of web pages to identify phishing websites\cite{visualmethod1}\cite{visualmethod2}. Chen el al \cite{visualmethod3} also proposed discriminative keypoint features that consider phishing detection as an image matching problem. Visual similarity-based approaches have their limitations, for example, methods that use the content of a website will fail to detect websites that use images instead of text. Methods that use image matching methods are very time-consuming and hard to gather enough data\cite{visualmethod4}.
\item \textit{Heuristics and machine learning based}: 
Machine learning methods have proved to be a powerful tool to classify malicious activities or artifacts like spam emails or phishing websites. Most of these methods require training data, fortunately, there are many phishing website samples to train a machine learning model. Some machine learning methods use vision techniques by analyzing a snapshot of a website\cite{rao2015computer} and some of them use content and features of the website for phishing detection. Multiple machine learning methods have been used to detect phishing websites some of which are Logistic regression, decision tree, random forest, Ada boost, SVM, KNN, neural networks, gradient boosting, and XGBoost which are described in the following section.
\end{enumerate}

In a recent study \cite{gupta2018defending} on phishing, the authors emphasized that when some new solutions were proposed to overcome various phishing attacks, attackers evolve their method to bypass the newly proposed phishing method. Therefore, the use of hybrid models and machine learning-based methods is highly recommended. In this paper, we are going to use machine learning-based classifiers for detecting phishing websites.

\begin{figure}[h]
	\centering
	\includegraphics[width=0.5\textwidth]{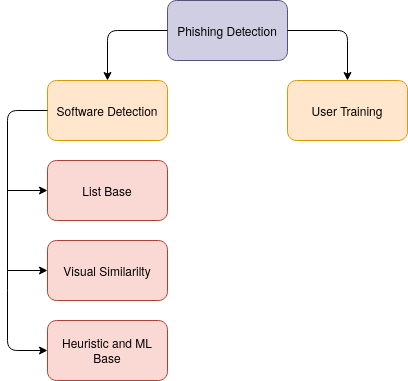}
	\caption{An Overview of phishing detection approaches}
	\label{fig:fig3}
\end{figure}

\section{MACHINE LEARNING APPROACH}\label{section-3}

Machine learning provides simplified and efficient methods for data analysis. It has indicated promising outcomes in real-time classification problems recently. The key advantage of machine learning is the ability to create flexible models for specific tasks like phishing detection. Since phishing is a classification problem, Machine learning models can be used as a powerful tool. Machine learning models could adapt to changes quickly to identify patterns of fraudulent transactions that help to develop a learning-based identification system.  Most of the machine learning models discussed here are classified as supervised machine learning, This is where an algorithm tries to learn a function that maps an input to an output based on example input-output pairs. It infers a function from labeled training data consisting of a set of training examples. We present machine learning methods that we used in our study.

\subsection{Logistic Regression}\label{AA}
Logistic Regression is a classification algorithm used to assign observations to a discrete set of classes. Unlike linear regression which outputs continuous number values, Logistic Regression transforms its output using the logistic sigmoid function to return a probability value which can then be mapped to two or more discrete classes. Logistic regression works well when the relationship in the data is almost linear despite if there are complex nonlinear relationships between variables, it has poor performance. Besides, it requires more statistical assumptions before using other techniques.

\subsection{K Near Neighbors} 
K-Nearest Neighbors (KNN) is one of the simplest algorithms used in machine learning for regression and classification problems which is non-parametric and lazy. In KNN there is no need for an assumption for the underlying data distribution. KNN algorithm uses feature similarity to predict the values of new datapoints which means that the new data point will be assigned a value based on how closely it matches the points in the training set. The similarity between records can be measured in many different ways. Once the neighbors are discovered, the summary prediction can be made by returning the most common outcome or taking the average. As such, KNN can be used for classification or regression problems. There is no model to speak of other than holding the entire training dataset. 

\subsection{Support Vector Machine}
Support vector machines (SVMs) are one of the most popular classifiers. The idea behind SVM is to get the closest point between two classes by using the maximum distance between classes. This technique is a supervised learning model used for linear and nonlinear classification. Nonlinear classification is performed using a kernel function to map the input to a higher-dimensional feature space. Although SVMs are very powerful and are commonly used in classification, it has some weakness. They need high calculations to train data. Also, they are sensitive to noisy data and are therefore prone to over-fitting. The four common kernel functions at the SVM are linear, RBF (radial basis function), sigmoid, and polynomial, which is listed in Table \ref{SVM-Kernel}. Each kernel function has particular parameters that must be optimized to obtain the best result.

\begin{table}[htbp]
	\caption{Four common kernels \cite{karatzoglou2006support}  }
	\begin{center}
		\begin{tabular}{ccc}
			\toprule
			\textbf{Kernel Type} & \textbf{Formula} & \textbf{Parameter} \\
			\midrule
			\textbf{Linear} & \(\displaystyle K(x_n,x_i) =(x_n,x_i) \)  & C,$\gamma$  \\ 
			\textbf{RBF} &  \(\displaystyle K(x_n,x_i) =exp(-\gamma\|x_n-x_i\|^2+C) \)  & C,$\gamma$ \\
			\textbf{Sigmoid} & \(\displaystyle K(x_n,x_i) =tanh(\gamma(x_n,x_i)+r) \)  & C,$\gamma$,r  \\
			\textbf{Polynomial} & \(\displaystyle K(x_n,x_i) =(\gamma(x_n,x_i)+r)^d \)  & C,$\gamma$,r,d \\ 
			\bottomrule
		\end{tabular}
		\label{SVM-Kernel}
	\end{center}
\end{table}

\subsection{Decision Tree}

Decision tree classifiers are used as a well-known classification technique. A decision tree is a flowchart-like tree structure where an internal node represents a feature or attribute, the branch represents a decision rule, and each leaf node represents the outcome. The topmost node in a decision tree is known as the root node. It learns to partition based on the attribute value. It partitions the tree in a recursive manner called recursive partitioning. This particular feature gives the tree classifier a higher resolution to deal with a variety of data sets, whether numerical or categorical data. Also, decision trees are ideal for dealing with nonlinear relationships between attributes and classes. Regularly, an impurity function is determined to assess the quality of the division for each node, and the Gini Variety Index is used as a known criterion for the total performance. In practice, the decision tree is flexible in the sense that it can easily model nonlinear or unconventional relationships. It can interpret the interaction between predictors. It can also be interpreted very well because of its binary structure. However, the decision tree has various drawbacks that tend to overuse data. Besides, updating a decision tree by new samples is difficult.

\subsection{Random Forest}
Random Forest, as its name implies, contains a large number of individual decision trees that act as a group to decide the output. Each tree in a random forest specifies the class prediction, and the result will be the most predicted class among the decision of trees. The reason for this amazing result from Random Forest is because of the trees protect each other from individual errors. Although some trees may predict the wrong answer, many other trees will rectify the final prediction, so as a group the trees can move in the right direction. Random Forests achieve a reduction in overfitting by combining many weak learners that underfit because they only utilize a subset of all training samples Random Forests can handle a large number of variables in a data set. Also, during the forest construction process, they make an unbiased estimate of the generalization error. Besides, they can estimate the lost data well. The main drawback of Random Forests is the lack of reproducibility because the process of forest construction is random. Besides, it is difficult to interpret the final model and subsequent results, because it involves many independent decision trees.\cite{breiman2001random}

\subsection{Ada-Boost}
From some aspects, Ada-boost is like Random Forest, the Ada-Boost classification like Random Forest groups weak classification models to form a strong classifier. A single model may poorly categorize objects. But if we combine several classifiers by selecting a set of samples in each iteration and assign enough weight to the final vote, it can be good for the overall classification. Trees are created sequentially as weak learners and correcting incorrectly predicted samples by assigning a larger weight to them after each round of prediction. The model is learning from previous errors. The final prediction is the weighted majority vote (or weighted median in case of regression problems). In short Ada-Boost algorithm is repeated by selecting the training set based on the accuracy of the previous training. The weight of each classifier trained in each iteration depends on the accuracy obtained from previous ones \cite{adaboost}.

\subsection{Gradeint Boosting}
Gradient Boosting trains many models incrementally and sequentially. The main difference between Ada-Boost and Gradient Boosting Algorithm is how algorithms identify the shortcomings of weak learners like decision trees. While the Ada-Boost model identifies the shortcomings by using high weight data points, Gradient Boosting performs the same methods by using gradients in the loss function. The loss function is a measure indicating how good the model’s coefficients are at fitting the underlying data. A logical understanding of loss function would depend on what we are trying to optimize.\cite{gardientBoosting}

\subsection{XGBoost}
XGBoost is a refined and customized version of a Gradient Boosting to provide better performance and speed. The most important factor behind the success of XGBoost is its scalability in all scenarios. The XGBoost runs more than ten times faster than popular solutions on a single machine and scales to billions of examples in distributed or memory-limited settings. The scalability of XGBoost is due to several important algorithmic optimizations. These innovations include a novel tree learning algorithm for handling sparse data; a theoretically justified weighted quantile sketch procedure enables handling instance weights in approximate tree learning. Parallel and distributed computing make learning faster which enables quicker model exploration. More importantly, XGBoost exploits out-of-core computation and enables data scientists to process hundreds of millions of examples on a desktop. Finally, it is even more exciting to combine these techniques to make an end-to-end system that scales to even larger data with the least amount of cluster resources.\cite{chen2016xgboost}

\subsection{Artificial Neural Networks}
\begin{figure}[h]
	\centering
	\includegraphics[width=0.5\textwidth]{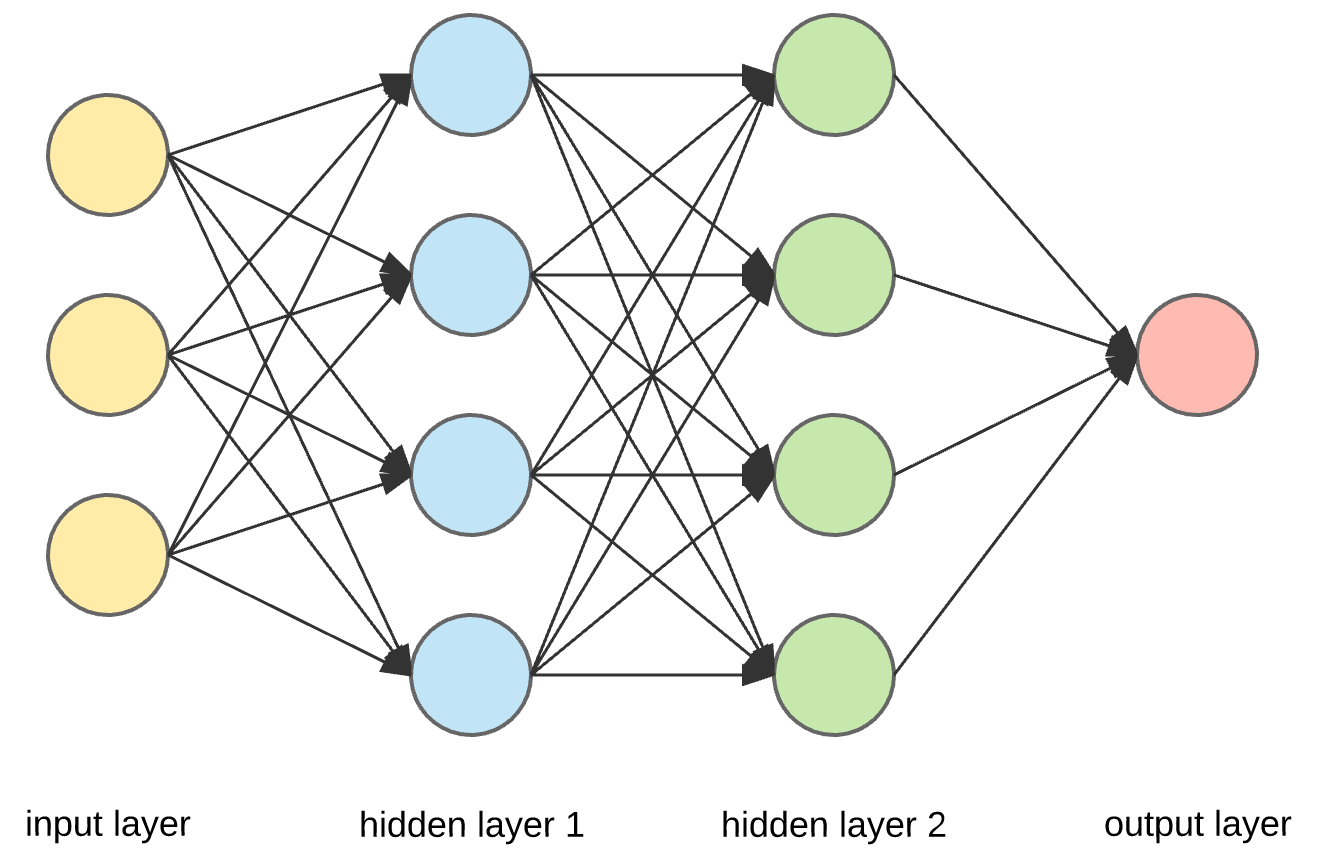}
	\caption{Artificial Neural Network}
	\label{fig:fig4}
\end{figure}

Artificial neural networks (ANNS) are a learning model roughly inspired by biological neural networks. These models are multilayered, each layer containing several processing units called neurons. Each neuron receives its input from its adjacent layers and computes its output with the help of its weight and a non-linear function called the activation function. In feed-forward neural networks like in \ref{fig:fig4}, data flows from the first layer to the last layer. Different layers may perform different transformations on their input. The weights of neurons are set randomly at the start of the training and they are gradually adjusted by the help of the gradient descent method to get close to the optimal solution. The power of neural networks is due to the non-linearity of hidden nodes. As a result, introducing non-linearity in the network is very important so that you can learn complex functions\cite{goodfellow2016deep}.

\section{Data set description}\label{section-4}
One of the main challenges in our research was the scarcity of phishing dataset. Although many scientific papers about phishing detection have been published, they have not provided the dataset on which they used in their research. Moreover, another factor that hinders finding a desirable dataset is the lack of a standard feature set to record characteristics of a phishing website. The dataset that we used in our research was well researched and benchmarked by some researchers. Fortunately, the accompanying wiki of the dataset comes with a data description document which discusses the data generation strategies taken by the authors of the dataset \cite{phishing-dataset}. For updating our dataset with new phishing websites we have also implemented a code that extracts features of new phishing websites that are provided by the PhishTank website. The dataset contains about 11,000 sample websites, we used 10\% of samples in the testing phase. Each website is marked either legitimate or phishing. The features of our dataset are as follows:

\begin{enumerate}
\item \textbf{Having IP Address}: If an IP address is used instead of the domain name in the URL, such as “http://217.102.24.235/sample.html”.
\item \textbf{URL Length}: Phishers can use a long URL to hide the doubtful part in the address bar.
\item \textbf{Shortening Service}: Links to the webpage that has a long URL. For example, the URL “http://sharif.hud.ac.uk/” can be shortened to “bit.ly/1sSEGTB”.
\item \textbf{Having @ Symbol}: Using the “@” symbol in the URL leads the browser to ignore everything preceding the “@” symbol and the real address often follows the “@” symbol
\item \textbf{Double Slash Redirection}: The existence of “//” within the URL which means that the user will be redirected to another website
\item \textbf{Prefix Suffix}: Phishers tend to add prefixes or suffixes separated by (-) to the domain name so that users feel that they are dealing with a legitimate webpage. For example http://www.Confirme-paypal.com.
\item \textbf{Having Sub Domain}: Having subdomain in URL.
\item \textbf{SSL State}: Shows that website use SSL
\item \textbf{Domain Registration Length}: Based on the fact that a phishing website lives for a short period
\item \textbf{Favicon}: A favicon is a graphic image (icon) associated with a specific webpage. If the favicon is loaded from a domain other than that shown in the address bar, then the webpage is likely to be considered a Phishing attempt.
\item \textbf{Using Non-Standard Port}: To control intrusions, it is much better to merely open ports that you need. Several firewalls, Proxy and Network Address Translation (NAT) servers will, by default, block all or most of the ports and only open the ones selected
\item\textbf{HTTPS token}: Having deceiving “https” token in URL. For example, “http://https-www-mellat-phish.ir”
\item \textbf{Request URL}: Request URL examines whether the external objects contained within a webpage such as images, videos, and sounds are loaded from another domain.
\item \textbf{URL of Anchor}: An anchor is an element defined by the $<a>$ tag. This feature is treated exactly as “Request URL”.
\item\textbf{Links In Tags}: It is common for legitimate websites to use <Meta> tags to offer metadata about the HTML document; <Script> tags to create a client side script; and <Link> tags to retrieve other web resources.
\item \textbf{Server Form Handler}: If the domain name in SFHs is different from the domain name of the webpage.
\item \textbf{Submitting Information To E-mail}: A phisher might redirect the user’s information to his email.
\item \textbf{Abnormal URL}: It is extracted from the WHOIS database. For a legitimate website, identity is typically part of its URL.
\item \textbf{Website Redirect Count}: If the redirection is more than four-time
\item \textbf{Status Bar Customization}: Use JavaScript to show a fake URL in the status bar to users
\item \textbf{Disabling Right Click}: It is treated exactly as “Using onMouseOver to hide the Link”
\item \textbf{Using Pop-up Window}: Showing having popo-up windows on the webpage.
\item \textbf{IFrame}: IFrame is an HTML tag used to display an additional webpage into one that is currently shown.
\item\textbf{Age of Domain}: If the age of the domain is less than a month.
\item\textbf{DNS Record}: Having the DNS record
\item \textbf{Web Traffic}: This feature measures the popularity of the website by determining the number of visitors.
\item \textbf{Page Rank}: Page rank is a value ranging from “0” to “1”. PageRank aims to measure how important a webpage is on the Internet.
\item \textbf{Google Index}: This feature examines whether a website is in Google’s index or not.
\item \textbf{Links Pointing To Page}: The number of links pointing to the web page.
\item \textbf{Statistical Report}: If the IP belongs to top phishing IPs or not.
\end{enumerate}

\begin{figure}[h]
	\centering
	\includegraphics[width=0.5\textwidth]{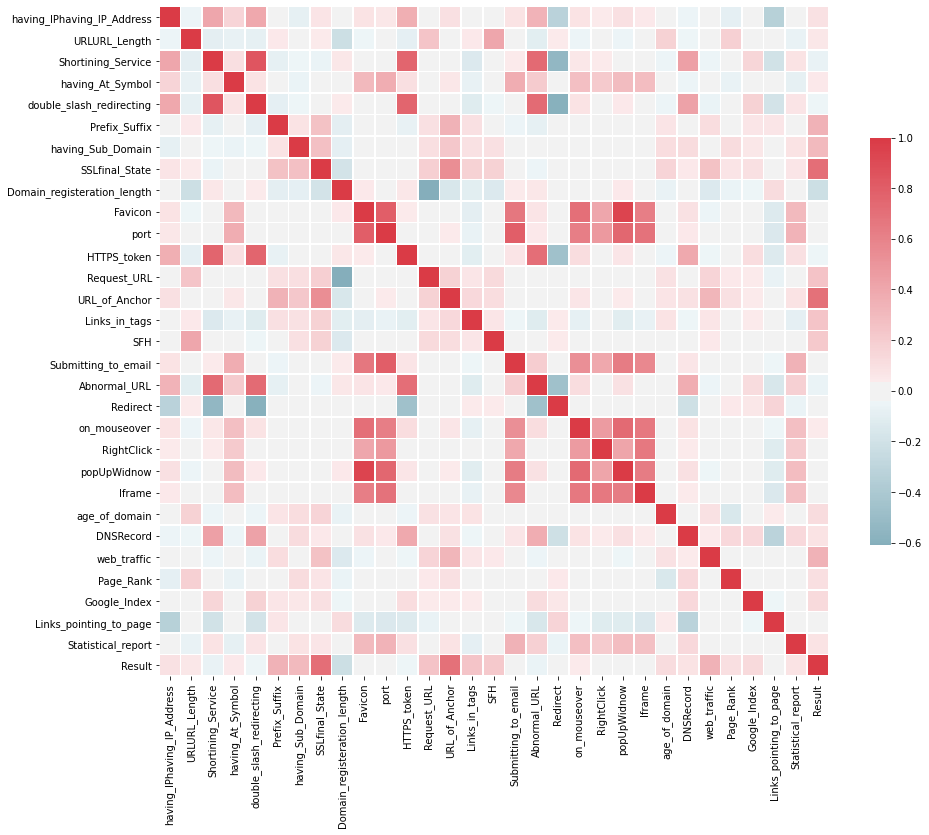}
	\caption{Corrolation of features in datasets}
	\label{fig:fig5}
\end{figure}

\begin{table}[htbp]
	\caption{Description of Dataset}
	\begin{center}
		\begin{tabular}{|c|c|c|}
			\hline
			\textbf{\textbf{features}} & \textbf{mean} & \textbf{std} \\
			\hline
			\textbf{Having IP Address} &  0.3137 &0.9495\\ 
			\hline
			\textbf{URL Length} & -0.6331 & 0.7660 \\ 
			\hline
			\textbf{Shortening Service} & 0.7387 & 0.6739 \\ 
			\hline
			\textbf{Having @ Symbol} & 0.7005 & 0.7135\\ 
			\hline
			\textbf{Double Slash Redirecting} &0.7414 &0.6710 \\ 
			\hline
			\textbf{Prefix Suffix} &-0.7349 &0.6781 \\ 
			\hline
			\textbf{Having Sub Domain} &0.0639 & 0.8175\\ 
			\hline
			\textbf{SSL Final State} &0.2509 & 0.9118\\ 
			\hline
			\textbf{Domain Reg Length	} & -0.3367& 0.9416\\ 
			\hline
			\textbf{Favicon} &0.6285 &0.7777 \\ 
			\hline
			\textbf{Port} & 0.7282&0.6853 \\ 
			\hline
			\textbf{HTTPS Token} &0.6750 &0.7377 \\ 
			\hline
			\textbf{Request URL} & 0.1867& 0.9824\\ 
			\hline
			\textbf{URL of Anchor} &-0.0765 &0.7151 \\ 
			\hline
			\textbf{Links in Tags} &-0.1181 &0.7639 \\ 
			\hline
			\textbf{SFH} &-0.5957 &0.7591 \\ 
			\hline
			\textbf{Submitting To Email} &0.6356 &0.7720 \\ 
			\hline
			\textbf{Abnormal URL} &0.7052 &0.7089 \\ 
			\hline
			\textbf{Website Redirect Count} &0.1156 &0.3198 \\ 
			\hline
			\textbf{On Mouse over} &0.7620 &0.6474 \\ 
			\hline
			\textbf{RightClick} & 0.9138&0.4059 \\ 
			\hline
			\textbf{PopUpWidnow} &0.6133 &0.7898\\ 
			\hline
			\textbf{IFrame} & 0.8169&0.5767 \\ 
			\hline
			\textbf{Age of Domain} & 0.0612& 0.9981\\ 
			\hline
			\textbf{DNS Record} &0.3771 & 0.9262\\ 
			\hline
			\textbf{Web Traffic} & 0.2872 & 0.8277 \\ 
			\hline
			\textbf{Page Rank} & -0.4836 &  0.8752\\ 
			\hline
			\textbf{Google Index} & 0.7215 & 0.6923 \\ 
			\hline
			\textbf{Links Pointing to Page} &0.3440 & 0.5699\\ 
			\hline
			\textbf{Statistical Report} & 0.7195& 0.6944\\ 
			\hline
			\textbf{Result} & 0.1138 &0.9935 \\ 
			\hline	
		\end{tabular}
		\label{tab2}
	\end{center}
\end{table}
\section{Evaluation Metrics}\label{section-5}
For evaluating phishing classification performance we use accuracy(acc) recall(r), precision(p), F1 score, test time, and train time of classifiers. Recall measures the percentage of phishing websites that the model manages to detect (model’s effectiveness). Precision measures the degree to which the phishing detected websites are indeed phishing (model’s safety). F1 score is the weighted harmonic mean of precision and recall. Let $N_{L \rightarrow L}$ be the number of legitimate websites classified as legitimate, $N_{L\rightarrow P}$ be the number of legitimate websites misclassified as phishing, $N_{P\rightarrow L}$ be the number of phishing misclassified as legitimate and $N_{P\rightarrow P}$  be the number of phishing websites classified as phishing. Thus the following equations hold
\begin{equation}
acc = \frac{N_{L\rightarrow L}+N_{P \rightarrow P}}{N_{L\rightarrow L} + N_{L\rightarrow P}+N_{P \rightarrow L}+N_{P \rightarrow P}} 
\end{equation}
\begin{equation}
r = \frac{N_{P \rightarrow P}}{N_{P \rightarrow L}+N_{P \rightarrow P}} 
\end{equation}
\begin{equation}
p = \frac{N_{P \rightarrow P}}{N_{L \rightarrow P}+N_{P \rightarrow P}} 
\end{equation}
\begin{equation}
F1 = \frac{2pr}{p+r} 
\end{equation}

\section{Experimental results}\label{section-6}
In our experiments, we used 10-fold cross-validation for model performance evaluation. we divided the data set into 10 sub-samples. A sub-sample is used for testing data and the rest is used for training models. Since phishing detection is a classification problem we must use a binary classification model, we consider ``-1`` as a phishing sample and ``1`` as a legitimate one.

\begin{table*}[htb]
	\caption{Classification Results for Different Methods}
\begin{tabularx}{\textwidth}{XXXXXXX}
	\toprule
	{\textbf{classifier}}    & \textbf{train time (s)} & \textbf{test time(s)}  & \textbf{accuracy} & \textbf{recall} & \textbf{precision} & \textbf{F1 score} \\
	\midrule
	{\textbf{logistic regression}} & 0.080971 & 0.006414 & 0.926550 & 0.943968 & 0.925700 & 0.934704 \\
	{\textbf{decision tree}} & 0.021452 & 0.003737 & 0.965988 & 0.971414 & 0.967681 & 0.969531 \\
	{\textbf{random forest}} & 0.436126 & 0.021941 & 0.972682 & 0.981484 & 0.969852 & 0.975622 \\
	{\textbf{ada booster}} & 0.336519 & 0.016766 & 0.936953 & 0.954362 & 0.933943&0.944032 \\
	{\textbf{KNN}} & 0.112972 & 0.353562 & 0.952780 & 0.962968 & 0.952783 & 0.957827 \\
	{\textbf{neural network}} & 9.088517 &	0.006925 & 0.969879 & 0.978723 & 0.967605 & 0.973112 \\
	\textbf{SVM\_linear} & 1.647538 & 0.053979 & 0.927726 & 0.945592 & 0.926268 & 0.935779  \\
	\textbf{SVM\_poly}  & 1.048257 &	0.074207 & 0.949254 & 0.968816 & 0.941779 &	0.955083 \\
	\textbf{SVM\_rbf}   & 1.341540 &	0.103329 & 0.952149 & 0.968815 & 0.946580 &	0.957543 \\
	\textbf{SVM\_sigmoid}    & 1.344607 &	0.109696 & 0.827498 & 0.846515 & 0.844311 &	0.845305 \\
	{\textbf{gradient boosting}} & 0.891888 & 0.005298 & 0.948621 & 0.962481 & 0.946234 & 0.954260 \\
	{\textbf{XGBoost}}  & 0.506072 & 0.006237 & 0.983235 & 0.981047 & 0.987235 & 0.976802 \\
	\bottomrule
\end{tabularx}
\label{tab3}
\end{table*}
In our study, we used various machine learning models for detection phishing websites which are Logistic regression, Ada booster, random forest, KNN, neural networks, SVM, Gradient boosting, XGBoost. We evaluate the accuracy, precision, recall, F1 score, training time, and testing time of these models and we used different methods of feature selection and hyperparameters tuning for getting the best results. Table \ref{tab2} shows the comparison between accuracy, precision, recall, and F1 score of these models.

For finding the best performance from support vector machine we have tested four kinds of kernel:
\begin{itemize}
	\item Linear kernel
	\item Polynomial kernel
	\item Sigmoid kernel
	\item RBF kernel
\end{itemize}

\begin{figure}[h]
	\centering
	\includegraphics[width=0.5\textwidth]{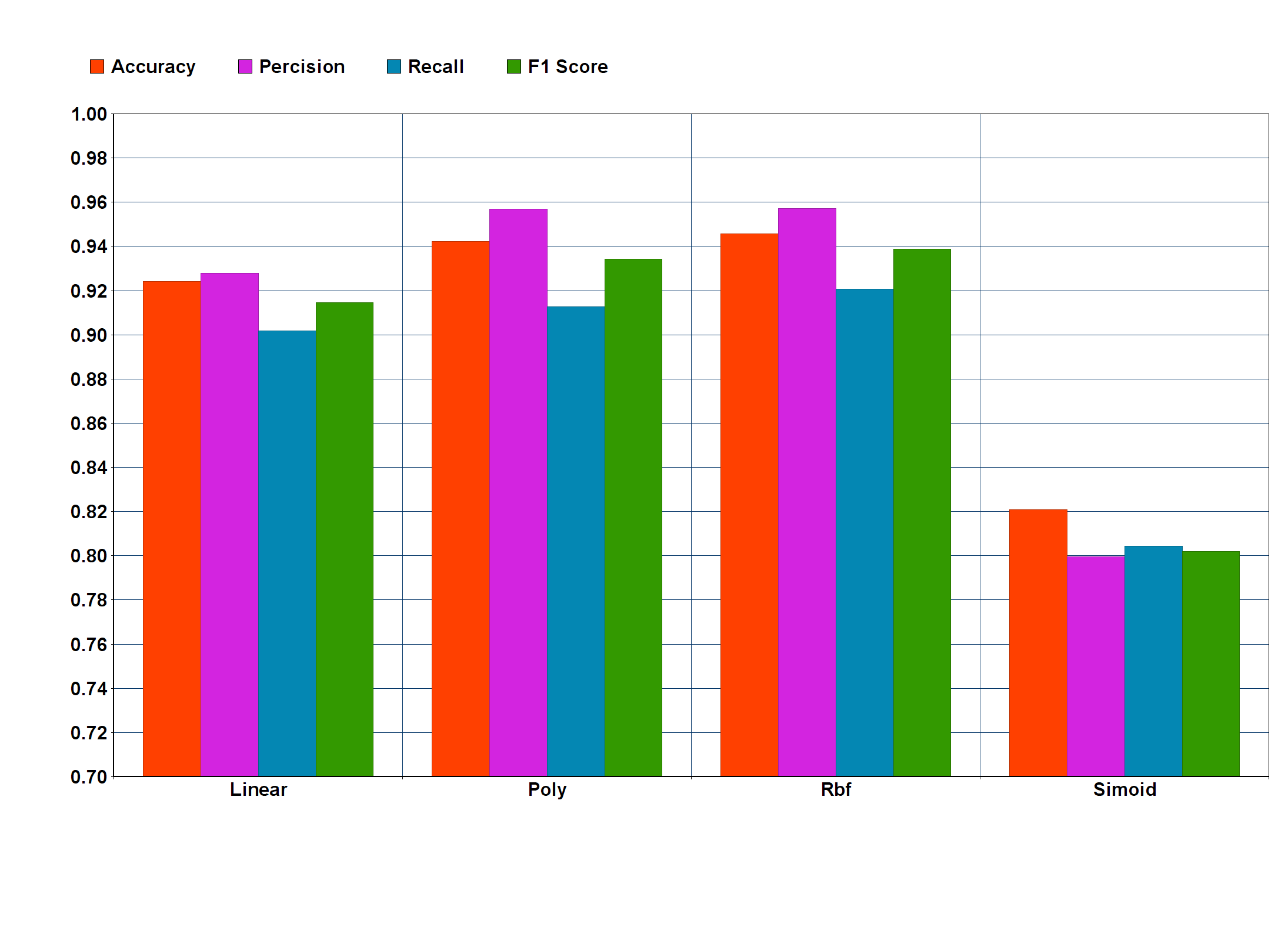}
	\caption{Performance of SVM classfier with various kernels}
	\label{fig:fig6}
\end{figure}

In our experience Linear, Polynomial, and RBF kernels would work equally well on this dataset but we get the best performance from the RBF kernel. The choice of the kernel and regularization parameters can be optimized with a cross-validation model selection. With more than a few hyper-parameters to tune, automated model selection is likely to result in severe over-fitting, due to the variance of the model selection criterion. In the absence of expert knowledge, the RBF kernel makes a good default kernel when our problem requiring a non-linear classifier. In Figure \ref{fig:fig6} performance of SVM with the different kernel are presented.

We found that Random Forest is highly accurate, relatively robust against noise and outliers, it is fast, simple to implement and understand, and can do feature selection implicitly. being unaffected by noise is the main advantage of Random Forest over AdaBoost. According to Central Limit Theorem, Random Forest reduces variance by increasing the number of trees. However, the main disadvantage of Random Forests that we faced in implementing our model was the high number of hyperparameters to tune for getting the best performance. Moreover, Random Forest introduces randomness into the training and testing data which is not suitable for all data sets.

\begin{figure}[h]
	\centering
	\includegraphics[width=0.5\textwidth]{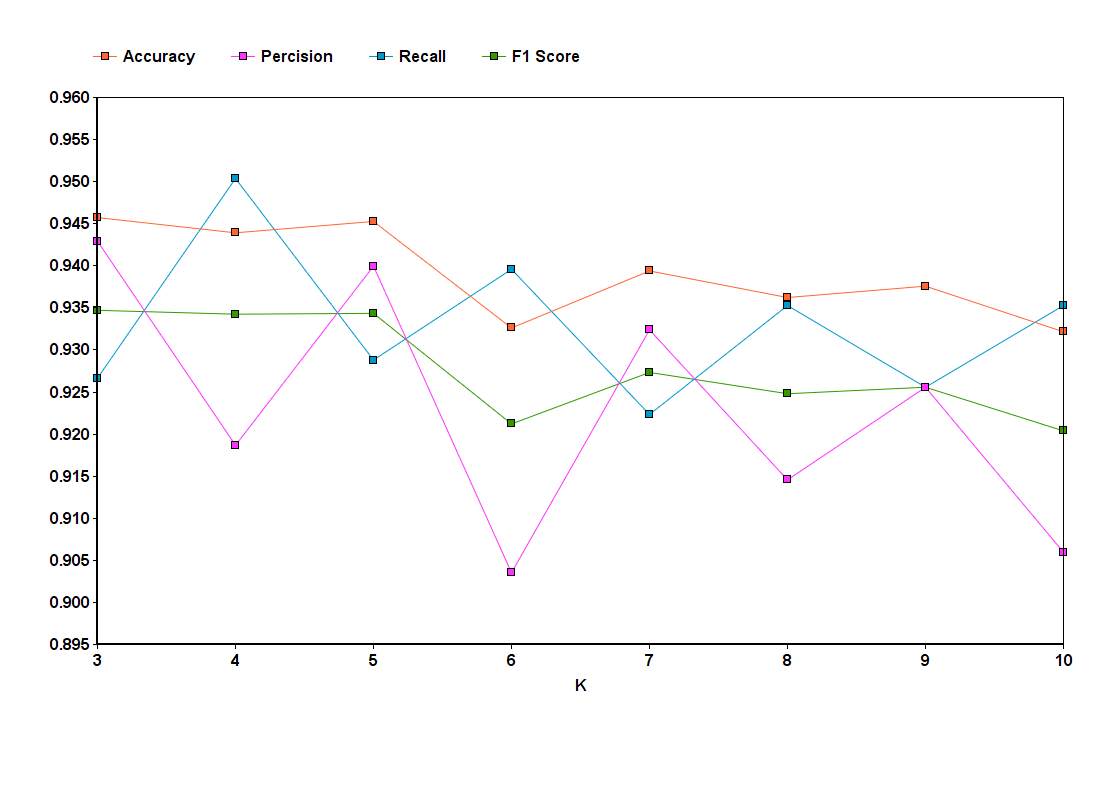}
	\caption{KNN with different K}
	\label{fig:fig7}
\end{figure}

In KNN classification we found out the best performance is acquired when we set k to 5. In KNN classification there is no optimal number to set k that is suitable for all kinds of datasets. According to the KNN result which is shown in Figure \ref{fig:fig7} the noise will have a higher impact on the result when the number of neighbors is small, moreover, a large number of neighbors make it computationally expensive to acquire the result. Our result has also shown that a small number of neighbors is the most flexible fit which will have low bias but the high variance plus a large number of neighbors will have a smoother decision boundary which means lower variance but higher bias.

The main advantage of XGBoost is its fast speed compared to other algorithms, such as ANN and SVM, and it's regularization parameter that successfully reduces variance. But even aside from the regularization parameter, this algorithm leverages a learning rate and subsamples from the features like random forests, which increases its ability to generalize even further. However, XGBoost is more difficult to understand, visualize, and to tune compared to AdaBoost and Random Forests. There is a multitude of hyperparameters that can be tuned to increase performance.XGBoost is a particularly interesting algorithm when speed as well as high accuracies are of the essence. Nevertheless, more resources in training the model are required because the model tuning needs more time and expertise from the user to achieve meaningful outcomes.

\begin{figure}[h]
	\centering
	\includegraphics[width=0.5\textwidth]{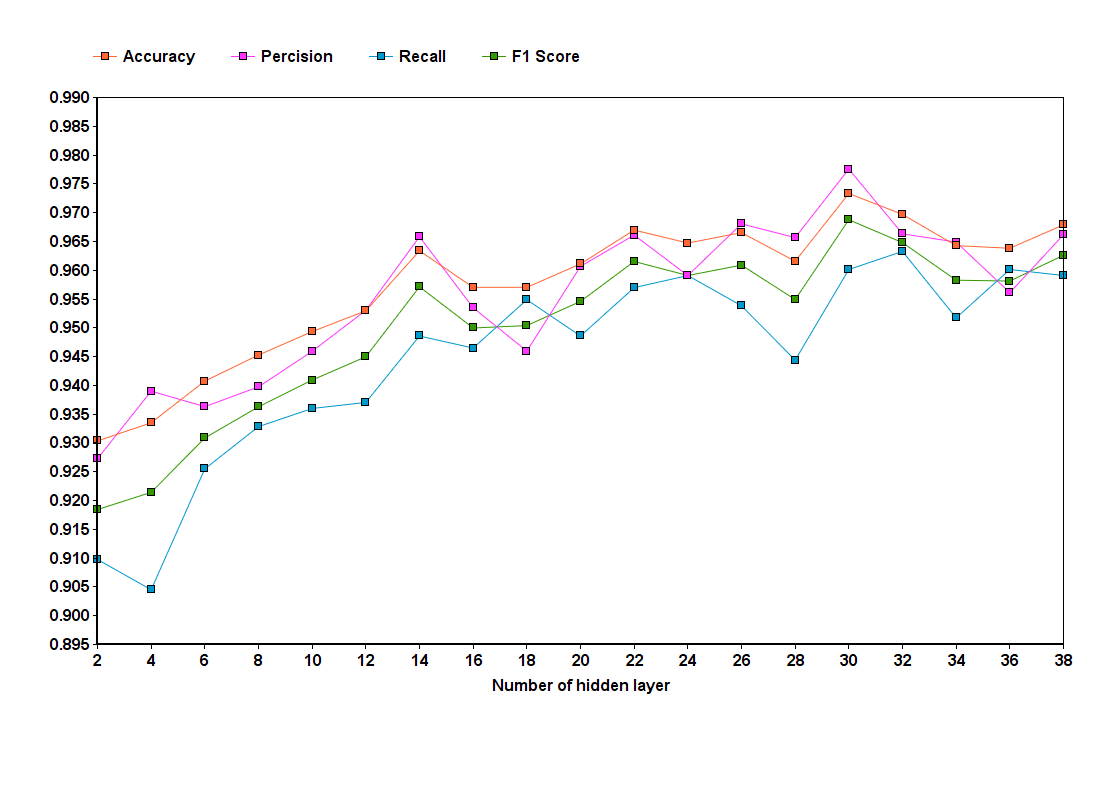}
	\caption{Neural Network with different depth}
	\label{fig:fig8}
\end{figure}

As expected, neural network's training time was considerably higher compared to other machine learning models. XGBoost's F1 score was slightly better compared with neural network's. This is due to the fact that our training data size is small. Unlike XGBoost, neural network model is also unable to explain why it have predicted a website as a phishing one. The explainability will help us to specify key features more easily. In the implementation of neural networks we use Adam optimizer and relu activation function in the hidden layer, figure \ref{fig:fig8} shows the performance of the neural network with a different number of the hidden layer, we get the best performance with 30 hidden layers. We trained our model on 500 epochs with early stopping.

\section{Conclusion and future work}\label{section-7}

In this research, we have implemented and evaluated twelve classifiers on the phishing website dataset that consists of 6157 legitimate websites and 4898 phishing websites. The examined classifiers are Logistic Regression, Decision Tree, Support Vector Machine, Ada Boost, Random Forest, Neural Networks, KNN, Gradient Boosting, and XGBoost. According to our result in Table \ref{tab3}, we get very good performance in ensembling classifiers namely, Random Forest, XGBoost both on computation duration and accuracy. The main idea behind ensemble algorithms is to combine several weak learners into a stronger one, this is perhaps the primary reason why ensemble-based learning is used in practice for most of the classification problems.
There are certain advantages and disadvantages inherent to the AdaBoost algorithm. AdaBoost is relatively robust to overfitting in low noisy datasets\cite{van2018hyperparameter}. AdaBoost has only a few hyperparameters that need to be tuned to improve model performance. Moreover, this algorithm is easy to understand and to visualize. However, for noisy data, the performance of AdaBoost is debated with some arguing that it generalizes well, while others show that noisy data leads to poor performance due to the algorithm spending too much time on learning extreme cases and skewing results. Compared to random forests and XGBoost, Moreover, AdaBoost is not optimized for speed, therefore being significantly slower than XGBoost.

It is worth mentioning that there is no guarantee that the combination of multiple classifiers will always perform better than the best individual classifier in the ensemble classifiers. The results motivate future works to add more features to the dataset, which could improve the performance of these models, hence it could combine machine learning models with other phishing detection techniques like example List-Base methods to obtain better performance. Besides, we will explore to propose and develop a new mechanism to extract new features from the website to keep up with new techniques in phishing attacks.

\section{Data and code}
To facilitate reproducibility of the research in this paper, all codes and data are shared at this GitHub repository : \href{https://github.com/fafal-abnir/phishing\_detection}{https://github.com/fafal-abnir/phishing\_detection}  

\section*{Acknowledgment}

This research was supported by Smart Land co. We would like to express our special thanks Abed Farvardin for providing us a resource for doing this project as well as Saeed Shahrivari who gave us the golden opportunity to do this wonderful project on the Phishing detection, which also helped us in doing a lot of research and we came to know about so many new things we are really thankful to them.

\bibliographystyle{ieeetr}
\bibliography{references}

\end{document}